\documentclass[12pt]{article}

\begin{document}

\title{Instanton constraints and renormalization}

\author{P. M. Glerfoss,  N.K. Nielsen
\footnote{Electronic address: nkn@fysik.sdu.dk}\\
 Physics Department, University of Southern Denmark,\\
Odense, Denmark.}

\date{\today}
\maketitle

\begin{abstract}
The renormalization is investigated of one-loop quantum fluctuations
around a constrained instanton in $\phi ^4$-theory with negative
coupling. It is found that the constraint should be renormalized
also. This indicates that in general only renormalizable constraints are
permitted.\\
PACS number 11.10.Ðz
\end{abstract}

\today

\section{Introduction} \label{Int}

Instantons \cite{Belavin}, \cite{Lipatov}  only exist in
conformally invariant field theories because of Derrick's
theorem \cite{Derrick}. When conformal invariance is
broken, approximate solutions can be used to estimate the path
integral of the theory in question \cite{'t Hooft};
one way of implementing this idea is to introduce a
constraint in the theory that explicitly violates conformal
invariance \cite{Frishman}, \cite{Affleck}.
 
In an earlier publication \cite{NN} it was shown that the
choice of constraint is more restricted than previously
assumed  since most constraints do not lead to a finite action.
Constrained instantons were explicitly constructed in two instances,
$\phi^4$-theory with "wrong" coupling sign, and the Yang-Mills-Higgs
theory. In the former case, where an instanton solution was first
obtained by Lipatov
\cite{Lipatov} in the context of the large-order
behaviour of the perturbation series, it was found that there
are only two permitted constraints, both involving a cubic
coupling, but either in operator form or as a source in the
field equation. For the Yang-Mills-Higgs case only a source constraint
was determined.

These results indicate that the allowed constraints should
have the property of being (i) renormalizable, (ii) gauge
invariant (in the case of gauge theories). In order to
investigate this conjecture further, we examine in the
present paper the quantum fluctuations around a constrained
instanton in $\phi ^4$-theory with wrong-signed coupling.
The Yang-Mills-Higgs system is more interesting physically,
but also more complicated. We compute the mass corrections of the
one-loop functional determinant, regularized by zeta function
regularization \cite{Hawking}, \cite{Elizalde}, and show that a change
of the mass scale introduced through regularization takes place through
the customary mass and coupling constant renormalization, if the
constraint is enforced by a source term, but with an operator
constraint also the constraint coefficient should be renormalized.
Thus it is indeed likely that the constraint always should be
renormalizable.

The layout of the paper is the following: In sec. 2 the instanton
solution of \cite{NN} is recapitulated, and the mass corrected terms
of the classical action relevant for renormalization are constructed; 
the construction involves an infinite resummation
because of infrared divergences. In sec. 3 a heuristic argument on the
three one-loop renormalizations of the theory in an instanton
background is given. Sec. 4 deals with the quantum fluctuations around
the instanton solution in the massless limit, and the mass corrections
of the eigenvalues are given in sec. 5. Finally in secs. 6 and 7 the
mass corrections of the one-loop functional determinant and their
renormalizations are dealt with for the cases of an operator and a
source constraint, respectively. Our results are briefly stated in the
conclusion, and three appendices deal with the eigenfunctions in the
massless case, matrix elements of the mass perturbation terms, and zeta
function regularization.  

\section{Instanton solution in massive $\phi ^4$-theory}

In the massless scalar $\phi^4$-theory with negative coupling constant the
Euclidean Lagrangian is
\begin{equation}
L(\phi )=\frac{1}{2}(\partial_\mu\phi)^2-\frac{g}{4!}\phi^4.
\end{equation}
The resulting equation of motion has an
instanton solution:
\begin{equation}
\phi_{0}=\sqrt{\frac{3}{g}}\frac{4\rho}{\rho ^2+x^2}
\label{inst}
\end{equation}
where the subscript indicates the order in the mass expansion and where
$\rho$ is an arbitrary scale parameter. The value of the action at the
instanton solution is:
\begin{eqnarray}
&&S[\phi_{0}]=\int d^4x L(\phi _0) =\frac{16\pi ^2}{g}.
\label{class}
\end{eqnarray}

In the presence of a mass term   one 
has to impose a constraint (that
also breaks scale invariance) in order to keep the action finite.
This may be a source constraint or an operator constraint. In the
massive scalar $\phi^4$-theory the appropriate operator constraint
is known to be \cite{NN}:
\begin{equation}
\int d^4x\phi^3(x)=k\rho
\label{konstraint}
\end{equation}
with $k$ a constant that to zeroth order has the value, found by
insertion of (\ref{inst}) into (\ref{konstraint}):
\begin{equation}
k_0=\frac{96}{g}\sqrt{\frac{3}{g}}\pi ^2.
\label{kvalg}
\end{equation}
The presence of the constraint leads to an
effective Lagrangian:
\begin{equation}
L_{\rm eff}(\phi )=L(\phi )+\frac{1}{2}m^2\phi^2
+\bar{\sigma }\phi ^3
\end{equation}
where $\bar{\sigma}$ is chosen such that the field equation
has a finite-action instanton solution. The mass parameter is assumed small so the
solution for $\bar{\sigma}$ and corrections to the instanton
configuration are expressed as a power series in $m$. 
To second order one finds the field equation: 
\begin{equation}
(\partial^2+\frac{g}{2}\phi _{0}^2)\phi _2 -m^2\phi
_{0}-3\bar{\sigma}_2\phi_{0}^2=0 \label{f¿rsteordens}
\end{equation}
with:
\begin{equation}
\bar{\sigma}_2=\frac{\sqrt{3g}}{6}m^2\rho
(\log\frac{m^2\rho^2}{4}+2\gamma+2).
\label{barrsigma}
\end{equation}
The corresponding  instanton
solution is \cite{NN}:
\begin{equation}
\phi_{\rm cl}\simeq \phi_{0}+\phi _{2}=\phi_{0}+\phi _{2, a}+
\phi _{2, b}, \label{insto}
\end{equation}
where
\begin{equation}
\phi _{2,a}=\sqrt{\frac{3}{g}}m^2\rho
(-(\frac{1}{1-u}-12u)\log u+6u(1-2u)\Phi
(\frac{1-u}{u})+12u),
\label{tiuve}
\end{equation}
with
\begin{equation}
u=\frac{\rho^2}{\rho
^2+x^2},\hspace{5mm}\Phi(x)=\int_0^xdx'\frac{\log(1+x')}{x'},
\end{equation}
and
\begin{equation}
\phi _{2, b}=\sqrt{\frac{3}{g}}m^2\rho
(\log\frac{m^2\rho^2}{4}+2\gamma-1).\label{instre}
\end{equation}
Here $\Phi (x)$ is the Spence function \cite{Mitchell}, and
$\phi _{2,b}$ constitutes with  $\phi _{0}$ (apart from a
proportionality factor) the two first terms of
the modified Bessel function
$K_1(m\mid x\mid )$, thus ensuring correct asymptotic
behaviour at   large values of
$\mid x\mid $, where
\begin{equation}
K_1(m\mid x\mid)\simeq \sqrt {\frac{\pi}{m\mid x\mid}}e^{-m\mid x\mid}.
\label{Bushel}
\end{equation}

To order $m^2$ the value of the constant $k$ is in fact
divergent, and  a large cutoff $R$ limiting
$\mid x\mid $ must be introduced. Then the following value of $k_2$ is
found:
\begin{eqnarray}&&
k_{2}
\simeq
\frac{144}{g}\sqrt{\frac{3}{g}}\pi ^2m^2\rho ^2
\Bigg( \frac{1}{2}\log ^2\frac{R^2}{\rho ^2}+\frac{\pi
^2}{6}+1
\nonumber\\&&
+
(\log\frac{m^2\rho^2}{4}+2\gamma-1)(\log \frac
{R^2}{\rho ^2}-1)\Bigg)
\label{f¿jitredie}
\end{eqnarray}
by means of (\ref{konstraint}) and the integrals (\ref{intlog}) and
(\ref{Folmer}) in App. \ref{int}.

The second-order mass corrections to the classical action are also
evaluated:
\begin{equation}
\frac{m^2}{2}\int d^4x \phi _{0}^2+\int d^4x(\partial _\mu
\phi _{0}
\partial _\mu \phi _{2}-\frac{g}{6}\phi _{0}^3\phi _{2}
).\label{massekorr}
\end{equation}
These integrals again diverge logarithmically but the divergences
actually cancel out after summation to all orders in the mass
parameter. Similar divergences also occur for the one-loop corrections
of the action. The resummation necessary to eliminate the divergences of
(\ref{massekorr}) is carried out by rewriting the action:
\begin{eqnarray}&&
S[\phi]=\int
d^4x(\frac{1}{2}\partial_\mu(\phi \partial_\mu \phi )+\frac{1}{2}\phi(-\partial
^2\phi-\frac{g}{6}\phi ^3+m^2\phi ^2+3\bar {\sigma }\phi ^2)
\nonumber\\&&+\frac{g}{24}\phi
^4-\frac{3}{2}\bar{\sigma }\phi ^3).
\label{resum}
\end{eqnarray}
Here the second term vanishes for a field configuration that is a solution of
the field equation. The first term  of (\ref{resum}):
\begin{equation}
\frac{1}{2}\int
d^4x\partial_\mu(\phi \partial_\mu \phi ) \label{surrface}
\end{equation}
is a surface term and is estimated by means of
the partition of the solution of the field equation into leading
terms, nextleading terms etc., where the
leading terms   for large $x$-values \cite{NN} sum to
\begin{equation}
4\rho \sqrt{\frac{3}{g}} \frac{m}{\mid x\mid}K_1(m\mid x\mid)
\end{equation}
with exponential falloff at large $\mid x\mid$.
(\ref{surrface}) thus goes to zero when the
integration volume goes to infinity.  The same conclusion
holds for nextleading etc. terms since they also after resummation have
exponential falloff \cite{NN} and the term (\ref{surrface}) can be
neglected altogether. Thus the value of the classical action at the
instanton solution $\phi _{cl}$ is:
\begin{equation}
S[\phi _{\rm cl}]=\int d^4x(\frac{g}{24}\phi_{\rm cl}
^4-\frac{3}{2}\bar{\sigma }\phi _{\rm cl}^3)
\end{equation}
that to second order is
\begin{equation}
S[\phi _{\rm cl}]\simeq \int
d^4x(\frac{g}{24}\phi _{0}^4+\frac{g}{6}\phi
_{0}^3\phi _{2, a}+\frac{1}{6}\phi _{0}^3(g\phi
_{2, b}-9\bar{\sigma })). \label{con}
\end{equation}

In (\ref{con}) the first term is known from  (\ref{class}), and the
third term is
\begin{eqnarray}&&
-\frac{24\pi ^2}{g}m^2\rho ^2(\log
\frac{m^2\rho ^2}{4}+2\gamma +8).
\end{eqnarray}
The second term of (\ref{con}) is by (\ref{inst}) and
(\ref{tiuve})  as well as (\ref{intlog}) and
(\ref{Folmer}):
\begin{equation}
\frac{g}{6}\int d^4x\phi _{0}^3\phi _{2, a} 
=\frac {168\pi ^2}{g} m^2\rho^2.
\end{equation}
The classical action including second order mass corrections is thus
\begin{equation}
S[\phi _{\rm cl}]\simeq \frac{16 \pi ^2}{g}(1-\frac {3m^2\rho
^2}{2}(\log \frac{m^2\rho ^2}{4}+2\gamma +1)).
\label{acsecond}
\end{equation}

Mass renormalization should, in the absence of infrared divergences,
be carried out upon $\frac{m^2}{2}\int\phi _0^2 d^4x$; after
resummation (\ref{acsecond}) should be used instead.
The coupling constant
renormalization  is not troubled by infrared divergences and should be
carried out on the mass corrected quartic part of the action before the
resummation leading to (\ref{acsecond}), i.e. on:
\begin{equation}
-\frac g6\int d^4x\phi _{2}\phi _{0}^3=-
\frac{48\pi ^2}{g}m^2\rho
^2(\log\frac{m^2\rho^2}{4}+2\gamma+\frac 52)
\label{phi4corr}
\end{equation}
while the  renormalization of the constraint coefficient, in the case of
an operator constraint, should act on:
\begin{equation}
\bar{\sigma }_2\int d^4x\phi _{0}^3=\bar{\sigma
}\frac{96}{g}\sqrt{\frac{3}{g}}\pi ^2\rho=\frac{48 \pi
^2}{g}m^2\rho ^2(\log
\frac{m^2\rho ^2}{4}+2\gamma +2)
\label{baresigma}
\end{equation}
in both cases by multiplication with the same coupling constant
renormalization factor.

\section{Quantum fluctuations}
\setcounter{equation}{0}
\subsection{One-loop path integral} \label{Eygordon}

The path integral is evaluated by the  Faddeev-Popov procedure,
modified to enforce the constraint. 
The outcome is in the one-loop approximation
\begin{equation}
Z\propto \int d\rho \int d^4z\mu (\rho 
)e^{-S[\phi _{\rm cl}]-\delta \bar{\sigma}_2\int
d^4x\phi _0^3}(\det{\hspace{0mm}}'M)^{-\frac 12}
\label{galoche}
\end{equation}
with integration over collective coordinates, where $\mu (\rho )$ is 
the appropriate integral measure, and with the Gaussian fluctuation
operator  
\begin{equation}
M\simeq M_0+M_2
\label{emma}
\end{equation}
with
\begin{equation}
M_0=\partial
^2-\frac{g}{2}\phi _{0}^2;\hspace{1mm}M_2=m^2-g\phi
_{2}\phi _{0}+6\bar{\sigma}_2\phi _{0}.
\end{equation}
Here the prime indicates that zero-modes and quasi-zero-modes; in the
massless limit there are four translational zero modes and one
dilatational zero mode, and after inclusion of mass they get nonzero
eigenvalues (the translational zero mode only if the constraint is
enforced by a source term). The integral measure $\mu (\rho )$ is 
 found from the normalization factors of the zero
modes and gets additional mass corrections from the nonzero
eigenvalue corrections of these modes. It will not be constructed
explicitly since it is irrelevant for the issue of renormalization. 

In (\ref{galoche}) the factor 
\begin{equation}
e^{-\delta \bar{\sigma}_2\int
d^4x\phi _0^3}, 
\label{persille}
\end{equation}
which only should be present for an operator
constraint, is generated by having a value of the constant
$k$ in the constraint (\ref{konstraint}), when applied to the
path integral, that is slightly different from $k_0$.
The factor (\ref{persille}) is necessary for the renormalization of the
constraint coefficient.

A normal coordinate expansion of the quantum fluctuation  scalar
field $\delta
\phi$ is:
\begin{equation}
\delta \phi=\sum a_p\phi _p \label{normabellini}
\end{equation}
where the sum runs over all  normalized field modes including the
translational zero modes $\phi _\mu$ and the dilational quasi-zero
mode $\phi _\rho$; $a_p$ denotes the corresponding normal
coordinates, and the eigenvalue equation is \cite{Lipatov}, \cite{'t
Hooft}:
\begin{equation}
M\phi _p=\lambda _p\frac{4\rho ^2}{(\rho ^2+x^2)^2}\phi
_p
\label{ayegen}
\end{equation}
where the nontrivial factor on the right-hand side makes the
spectrum discrete and reflects the fact that it conveniently may be
obtained by a stereographic projection \cite{Lipatov}.    Here the
eigenvalue
$\lambda _p$ is dimensionless, and since $\rho $ sets the instanton
scale, the proper dimensionful eigenvalue is $\frac{\lambda _p}{\rho
^2}$.

In the eigenvalue equation (\ref{ayegen}) the term $M_2$ should be
treated as a perturbation, and the equation can then be solved by
standard perturbation theory. However, at large distances a direct
determination of $\phi _p$ to all orders is possible. At large values of
$x^2$ the right-hand side (\ref{ayegen}) vanishes, and it reduces in
this limit
 to the Euclidean Klein-Gordon equation:
\begin{equation}
(-\partial^2+m^2)\phi _p\simeq 0
\end{equation}
with the normalizable solution 
\begin{equation}
\phi _p\propto\frac{m}{\mid x\mid }K_1(m\mid x\mid)
\label{croccodile}
\end{equation}
where the proportionality factor is fixed by comparison with the
massless limit.

\subsection{Green's function and renormalization} \label{Green}

A rough estimate of the mass correction of the functional
determinant, where only terms proportional to $\log (m^2\rho ^2)$ are
considered, is first given by means of Green's function techniques.
From this estimate a heuristic argument on the required
renormalizations is made. 

Starting from the free
massless propagator
\begin{equation}
D(x)=\frac{1}{4\pi ^2x^2}
\end{equation}
the determinant is computed perturbatively, with the perturbation
\begin{equation}
\Delta M= -\frac g2 \phi _{0}^2+
M_{2}
\end{equation}
where $M_{2}$ is defined in (\ref{emma}).
The massless propagator is then approximately
\begin{eqnarray}&&
G_{0}(x, x')
\simeq D(x-x')+\frac g2 \phi _{0}^2(x) \frac{1}{16\pi
^2}\log
\frac{R^2}{(x-x')^2}
\label{f¿rstegreen}
\end{eqnarray}
where $R$ is the infrared cutoff introduced previously.

The mass correction to the
one-loop action is then, also approximately:
\begin{equation}
\frac 12\int d^4 xM_{2}(x)G_{0}(x, x)
\simeq \frac g4\frac{1}{16\pi ^2 }\log (\Lambda ^2R^2)\int
d^4 xM_{2}(x)\phi_{0}^2(x)
\end{equation}
with $\Lambda $ an ultraviolet cutoff.  This expression has
a contribution from the term $m^2$ of (\ref{emma}):
\begin{eqnarray}&&
\frac g4 m^2\frac{1}{16\pi ^2 }\log (\Lambda ^2R^2)\int
d^4 x\phi_0^2(x) \simeq \frac 34 m^2\rho
^2\log (\Lambda^2R^2)\log \frac{R^2}{\rho ^2}
\nonumber\\&&
\simeq -\frac 34 m^2\rho
^2\log (\Lambda^2\rho^2)\log (m^2\rho ^2)
\label{r¿meren}
\end{eqnarray}
a contribution from $-g\phi _0\phi _2$
\begin{eqnarray}&&
-\frac {g^2}{4}\frac{1}{16\pi ^2 }\log (\Lambda ^2R^2)\int
d^4 x\phi_2(x)\phi_0^3(x) 
\simeq
-\frac 92m^2\rho ^2 \log (\Lambda ^2R^2)\log
(m^2\rho ^2)
\nonumber\\&&
\simeq
-\frac 92m^2\rho ^2 \log (\Lambda^2\rho^2)\log 
(m^2\rho ^2)
\label{r¿merto}
\end{eqnarray}
and a term from $6\bar{\sigma}_2\phi _0$:
\begin{eqnarray}&&
\frac g4\frac{1}{16\pi ^2 }\log (\Lambda
^2R^2)6\bar{\sigma }_2\int d^4 x\phi_0^3(x)
\simeq \frac 92m^2\rho ^2 \log (\Lambda ^2R^2)\log
(m^2\rho ^2)
\nonumber\\&&
\simeq \frac 92m^2\rho ^2 \log (\Lambda^2\rho^2)\log (m^2\rho ^2).
\label{r¿mertre}
\end{eqnarray}
In (\ref{r¿meren}), (\ref{r¿merto}) and (\ref{r¿mertre}) the infrared
cutoff
$\frac{1}{R}$ was replaced by either the instanton scale
parameter $\rho$ or the physical mass
$m$ in an apparently arbitrary manner; this procedure is justified
below in sec.
\ref{synapsid} and sec.
\ref{SourceC}. The two expressions in (\ref{r¿merto}) and
(\ref{r¿mertre}) actually cancel, but it is instructive to consider
them separately, since they correspond to different renormalizations.

The coupling constant renormalization is easily obtained by 
this argument also; going to second order
in (\ref{f¿rstegreen}) one gets the second-order correction to the
determinant:
\begin{equation}
-\frac{g^2}{16}\frac{1}{16\pi ^2 }\log (\Lambda ^2R^2)\int
d^4 x\phi_0^4 \simeq -\frac 32 \log (\Lambda ^2R^2).
\label{couplingling}
\end{equation}

 The divergent
expression (\ref{r¿meren}) is eliminated by the customary mass
renormalization, which in perturbation theory to lowest order is:
\begin{equation}
m^2_{\rm bare}\simeq m^2(1-\frac
{g}{32\pi ^2}\log\frac{\Lambda ^2}{\mu^2} )\label{massren} 
\end{equation}
with $\mu $ an arbitrary mass scale (only the logarithmic part is kept).

Replacing in (\ref{acsecond}) the mass with the bare mass according
to (\ref{massren}) one obtains the double logarithmic term
\begin{equation}
\frac 34m^2\rho ^2\log \frac {m^2\rho ^2}{4}\log \frac
{\Lambda ^2}{\mu^2}
\end{equation}
which exactly cancels the divergence in (\ref{r¿meren}).

For the expression (\ref{r¿merto})  the infinity is
removed by the coupling constant renormalization   of ordinary perturbation
theory of
(\ref{phi4corr}):
\begin{equation}
g_{\rm bare}\simeq g(1-\frac{3g}{32\pi ^2}\log \frac{\Lambda
^2}{\mu^2}).
\label{coupren}
\end{equation}
Notice that the renormalization is carried
out on the coupling in  the first version of (\ref{phi4corr}) where
$g$ is in the numerator.

Finally for (\ref{r¿mertre}) the infinity is removed by  replacing
$\bar{\sigma}_2$ in (\ref{baresigma}) by the corresponding bare
quantity:  
\begin{equation}
\bar{\sigma }_{ \rm bare ,2}\simeq \bar{\sigma
}_2(1-\frac{3g}{32\pi ^2}\log \frac{\Lambda ^2}{\mu^2})
\label{rubik}
\end{equation}
i.e. the cubic coupling is renormalized as the quartic
coupling, as in ordinary perturbation theory. In (\ref{persille}) this
corresponds to the choice:
\begin{equation}
\delta \bar{\sigma}_2=-\bar{\sigma}_2\frac{3g}{32\pi ^2}\log
\frac{\Lambda ^2}{\mu^2}.
\end{equation}

This loose argument indicates that the cubic operator constraint
should be renormalized as an ordinary cubic coupling; that this is
indeed the case is verified in detail below in secs. 6 and 7. With a
source constraint (\ref{r¿mertre}) is absent and only mass and
coupling constant renormalizations are required.

\section{The massless limit}
\setcounter{equation}{0}

In the massless limit the eigenvalue equation
(\ref{ayegen}) is explicitly solvable. The eigenfunctions $\phi _p$ 
are separated into a radial and an angular part:
\begin{equation} 
\phi _p(x)=\phi _{nlm_1m_2}(x)=(\sqrt
2\rho)^{-1}P_{lm_1m_2}(\Omega )u^{l+1}(1-u)^l\chi_{nl}(u)
\label{femtini}
\end{equation}
where the
angular part is an $O(4)$ spherical
harmonic
$P_{lm_1m_2}(\Omega )$
while the radial function $\chi_{nl}(u)$  can be expressed in terms of
a Gegenbauer polynomial; it is given explicitly in (\ref{chichi}) and
dealt with in detail in App. \ref{Gegencobi}.
The
angular eigenfunctions $P_{lm_1m_2}(\Omega )$ are normalized
on the unit three-sphere:
\begin{equation}
\int_{S^3}d\Omega P_{lm_1m_2}P_{l'm_1'm_2'}=\delta_{ll'}\delta_{m_1m_1'}
\delta_{m_2m_2'}.
\label{ofire}
\end{equation}
The  eigenfunctions are required to be normalized
according to:
\begin{eqnarray}
\int d^4x\frac{4\rho ^2}{(x^2+\rho ^2)^2}\phi_{n
lm_1m_2}\phi_{n' l'm_1'm_2'}=\delta _{ll'}\delta _{nn'}\delta
_{m_1m_1'}\delta _{m_2m_2'}.
\label{norm}
\end{eqnarray}
The eigenvalues are
\begin{equation}
\lambda_{nl}=(n+2l+4)(n+2l-1).
\label{egenv¾rdi}
\end{equation}
The degeneracy is $(2l+1)^2$.
The corresponding eigenvalues in the absence of an instanton are
\begin{equation}
\lambda_{nl}\mid _{\rm free}=(n+2l+1)(n+2l+2).
\end{equation}
The eigenfunctions are unmodified.
The functional determinant in the massless limit was computed in
Lipatov's original paper \cite{Lipatov} from (\ref{egenv¾rdi}).

There are five zero modes. The translational zero
modes have
$n=0$, $l=\frac{1}{2}$ and the dilatation
zero mode has $n=1$, $l=0$.
For $n=l=0$  an unstable mode occurs. The
existence of the unstable mode can be inferred already from
the zeroth order field equation  which is reformulated
\begin{equation}
-(\partial^2+\frac g2\phi _{0}^2)\phi
_{0}=-\frac{16\rho ^2}{(\rho ^2+x^2)^2}\phi _{0}.
\end{equation}
The eigenfunction of the unstable mode is thus proportional to
$\phi _{0}$.
The value of $\lambda _{00}$ can as shown by Lipatov
\cite{Lipatov} in the context of the estimate of large-order
perturbation theory can be taken as
$4$ instead of
$-4$.

\section{Mass corrections of the eigenvalues}

\setcounter{equation}{0}

\subsection{Statement of the problem}

The lowest-order mass corrections to the
eigenvalues found in the previous section are now computed. 
The perturbed Gaussian operator is given in (\ref{emma}).
The perturbed eigenvalue problem
reduces  to first order in $m^2$ to
\begin{equation}
M_{0}\phi_{nl,2}^{}+M_{2}\phi_{nl,0}
=\frac{4\rho ^2}{(\rho^2+x^2)^2}(\lambda_{nl,0}\phi_{nl,2}
+\lambda_{nl,2}\phi_{nl,0})
\label{puertuerb}
\end{equation}
whence: 
\begin{equation}
\lambda_{nl,2}=-\int d^4x\partial
_\mu(\phi_{nl,0}\stackrel{\leftrightarrow}{\partial
_\mu}\phi_{nl,2})+\int d^4x
M_{2}\phi_{nl,0}^2 \label{overflade}
\end{equation}
for eigenfunctions $\phi_{nl,0}$ normalized
according to (\ref{norm}). Here we have suppressed the quantum
numbers $m_1$ and $m_2$ but indicated the order in the mass
expansion for the eigenfunctions. The first term in (\ref{overflade}) is
a surface term at infinity that is nonvanishing if
$\phi_{nl,0}$ goes as
$\frac{1}{x^2}$ and
$\phi_{nl,2}$ has a constant term and a term that grows
logarithmically; this is the case for all
$l=0$-modes including the dilatational quasi-zero mode.

\subsection{The mass term correction}

With $M_2\rightarrow m^2$
the second term of (\ref{overflade}) is by (\ref{enoveruianden}):
\begin{eqnarray}&&
\int d^4x m^2\phi^{2}_{nl,0}
=\frac{m^2\rho ^2}{4}<nl\mid \mid \frac{1}{u^2}\mid \mid nl>
\nonumber\\&&
=\frac{m^2\rho
^2}{4}\frac{2n+4l+3}{2(2l+1)}\frac{(n+2l+1)(n+2l+2)}{l(l+1)}
\label{Solbakken}
\end{eqnarray}
where  the notation is explained in connection with (\ref{chichichi}). 

For $l=0$ a  logarithmic divergence  occurs in this matrix element.
This is seen by use of  (\ref{l0}), with the
integration volume a large sphere with radius  $R$: 
\begin{eqnarray}
&&\frac{m^2\rho ^2}{4}<n0\mid \mid u^{-2}\mid \mid n0>
\nonumber\\&&
=\frac{m^2\rho^2}{4}(2n+3)(1+(n+1)(n+2)(\log
\frac{R^2}{\rho ^2}-\frac 32-2\sum _{k=1}^n\frac{1}{k+1}))
\label{170}
\end{eqnarray}
that diverges logarithmically at  $R\rightarrow \infty $.
The infrared divergence is eliminated by using 
as a starting point the asymptotic expression for the
$m=0$ eigenfunction at large values of $x^2$ found from
(\ref{femtini}):
\begin{equation}
\phi_{n0,0}\simeq \sqrt{\frac{(2n+3)\Gamma (n+3)}{4\pi ^2n!}}
\frac{\rho }{x^2} \label{ras}
\end{equation}
whence is obtained the following asymptotic expression of the mass corrected
eigenfunction necessary to obtain the correct Bessel function according
to (\ref{croccodile}):
\begin{equation}
\phi _{n0,2}\simeq \frac 14 \rho m^2\sqrt{\frac{(2n+3)\Gamma
(n+3)}{4\pi ^2n!}} (\log\frac{m^2x^2}{4}+2\gamma-1). \label{deltaras}
\end{equation}
This leads to a correction to the eigenvalue from
the surface term of (\ref{overflade}) with the value
\begin{equation}
-\frac{m^2\rho ^2}{4}(2n+3)(n+1)(n+2)(\log \frac
{m^2R^2}{4}+2\gamma ).
\end{equation}
Adding this expression to (\ref{170}) one finds:
\begin{equation}
-\frac{m^2\rho^2}{4}(2n+3)(-1+(n+1)(n+2)(\log
\frac{m^2\rho ^2}{4}+2\gamma +\frac 32+2\sum
_{k=1}^n\frac{1}{k+1}))
\label{175}
\end{equation}
where the infrared divergences have cancelled as expected.

In the free-field case (absence of instanton) one has
\begin{equation}
M_{2}\rightarrow  M_{2, {\rm free}}=m^2
\end{equation}
and the eigenfunctions in the massless limit are the same as in the
presence of an instanton. 
Thus the total  eigenvalue corrections are in this case 
(\ref{Solbakken}) for $l\neq 0$ 
and (\ref{175}) for $l=0$.

\subsection{The remaining matrix element}

The last term of (\ref{overflade}) has the remaining term:
\begin{eqnarray}
&&\int d^4x
(M_{2}-m^2)\phi^{2}_{nl,0}
=\frac{m^2\rho ^2}{4}<nl\mid \mid
\frac{M_{2}-m^2}{m^2u^2}\mid \mid nl>
\nonumber\\&&
=\frac{m^2\rho ^2}{4}<nl\mid \mid (-12
u^{-1}[-(\frac{1}{1-u}-12u)\log u
\nonumber\\&&
+6u(1-2u)\Phi
(\frac{1-u}{u})+12u-3] )\mid \mid nl>.
\label{matrixelement}
\end{eqnarray}
The
matrix elements are found in Appendix
\ref{int} by rather laborious calculations, and the relevant results
are given  in (\ref{uminusen}), 
 (\ref{enelog}), (\ref{hunafloi}) and (\ref{looklook}), leading to
the result:
\begin{eqnarray}
&&\int d^4x
(M_{2}-m^2)\phi^2_{nl,0}
\nonumber\\&&
=-3m^2\rho
^2\frac{2n+4l+3}{2l+1}(2\sum
_{k=1}^{n+2l+1}\frac{1}{k+2l+1}+\frac{1}{2l+1}-3)
\nonumber\\&&
+18m^2\rho
^2(\frac{2l+1}{n+2l+2}\sum
_{k=0}^{n+2l+1}\frac{1}{k+2l+2}+\frac{2l+1}{n+2l+1}\sum_{k=0}^{n+2l}\frac{1}{k+2l+1}
\nonumber\\&&
+\frac{1}{n+4l+3}-\frac{1}{n+4l+2} -2).
\label{maatrixelement}
\end{eqnarray}

The sum of (\ref{Solbakken}) and (\ref{maatrixelement}) reduces to 0 in
the case
$n=0,l=\frac 12$, which corresponds to the translational zero modes.
This is expected, since the constraint is an operator constraint and
thus respects translational invariance.
For the dilatational zero mode with $n=1,l=0$
the eigenvalue correction is:
\begin{equation}
-\frac{15m^2\rho^2}{2}(\log
\frac{m^2\rho ^2}{4}+2\gamma +4)\neq 0.
\label{neeman}
\end{equation}
The dilatational zero mode thus becomes a quasi-zero mode
because of quantum fluctuations.

\section{Mass correction of functional determinant} \label{synapsid}

\setcounter{equation}{0}

The mass corrected functional determinant in (\ref{galoche}) is:
\begin{equation}
(\det (M_{0}+
M_{2}))^{-\frac12}\simeq(\det
M_{0})^{-\frac12}e^{-\frac 12{\rm tr}\frac{M_2}{M_0}}
\label{detto}
\end{equation}
where
\begin{eqnarray}&&
\frac 12{\rm tr}\frac{M_2}{M_0}=\frac
12\sum_{n=0}^\infty\sum_{l=0, n+2l>1}^\infty(2l+1)^2
\frac{<nl\mid \mid M_{2}\mid \mid
nl>}{(n+2l+4)(n+2l-1)}
\nonumber\\&&
-\frac 12\sum_{n=0}^\infty\sum_{l=0}^\infty(2l+1)^2
\frac{<nl\mid \mid M_{2, {\rm free}}\mid \mid
nl>}{(n+2l+1)(n+2l+2)}
\label{logdetto}
\end{eqnarray}
thus is the mass correction to the one-loop action.
Here the free-field contribution was subtracted. Also the unstable made
is disregarded here and henceforth since a single mode does not affect
the renormalizations.

The contribution to (\ref{logdetto}) from  the mass term of $M_2$ is
according to (\ref{Solbakken}) for $l\neq 0$:
\begin{eqnarray}
&&\frac{m^2\rho ^2}{16}(\sum_{n=0}^\infty\sum_{l=\frac 12,
n+2l>1}^\infty\frac{(2l+1)(n+2l+1)(n+2l+2)(2n+4l+3)}{(n+2l+4)(n+2l-1)l(l+1)}
\nonumber\\&&
-\sum_{n=0}^\infty\sum_{l=\frac 12}^\infty
\frac{(2l+1)(2n+4l+3)}{l(l+1)})
\label{enninul}
\end{eqnarray}
This expression is not well defined at $l=0$ where it according
to  (\ref{175})  is replaced by:
\begin{eqnarray}&&
-\frac{m^2\rho^2}{8}\sum _{n=2}^\infty 
\frac {2n+3}{(n+4)(n-1)}
(-1+(n+1)(n+2)(\log
\frac{m^2\rho ^2}{4}+2\gamma +\frac 32+2\sum
_{k=1}^n\frac{1}{k+1}))
\nonumber\\&&
+\frac{m^2\rho^2}{8}\sum _{n=0}^\infty \frac
{2n+3}{(n+1)(n+2)}(-1+(n+1)(n+2)(\log
\frac{m^2\rho ^2}{4}+2\gamma +\frac 32+2\sum
_{k=1}^n\frac{1}{k+1})). \label{183}
\nonumber\\&&
\end{eqnarray}
Introducing the summation variable $s=n+2l+\frac 32$
and adding (\ref{enninul}) and (\ref{183})  one gets:
\begin{equation}
-\frac{m^2\rho^2}{4}(B_\frac 52-B_\frac 12)
\label{seksogfirs}
\end{equation}
with
\begin{equation}
B_\phi =\sum _{s=\phi +1}^\infty
\frac {s}{s^2-\phi ^2}(-2+(s^2-\frac14)(\log
\frac{m^2\rho ^2}{4}+2\gamma +2)).
\label{befi}
\end{equation}

The function $B_\phi $ is zeta  function regulated \cite{Hawking}
\cite{Elizalde} in terms of the function $Z_\phi(\epsilon)$
defined by
\begin{equation} Z_\phi(\epsilon)=\sum _{s=\phi +1 }^\infty s
(s^2-\phi ^2)^{-\epsilon}. \label{Zeta}
\end{equation}
Replacing in (\ref{befi}) the denominator with the same quantity to the
power $1+\epsilon $ one gets:
\begin{eqnarray}&&
B^\epsilon _\phi
=(\log
\frac{m^2\rho ^2}{4}+2\gamma +2)Z_\phi(\epsilon )
\nonumber\\&&
+(-2+(\phi
^2-\frac14)(\log
\frac{m^2\rho ^2}{4}+2\gamma +2))Z_\phi(1+\epsilon ).
\end{eqnarray}
$B_\phi $ itself is determined from this expression and
(\ref{seksnulsyv}):
\begin{eqnarray}&&
B _\phi=(\log
\frac{m^2\rho ^2}{4}+2\gamma +2)Z_\phi(0) 
\nonumber\\&&
+(-2+(\phi
^2-\frac14)(\log
\frac{m^2\rho ^2}{4}+2\gamma +2))
\lim _{\epsilon \rightarrow 0}(\log
(\mu ^2\rho ^2)\epsilon Z_\phi(1+\epsilon )
+\frac{\partial }{\partial \epsilon}\epsilon Z_\phi(1+\epsilon ))
\nonumber\\&&
\label{btoreg}
\end{eqnarray}
that by (\ref{zetminusen}) and (\ref{zetanen}) is
\begin{eqnarray}&&
B _\phi=-(\log
\frac{m^2\rho ^2}{4}+2\gamma +2)(\frac
{1}{12}+\frac{\phi }{2})
\nonumber\\&&+(-1+\frac 12(\phi
^2-\frac14)(\log
\frac{m^2\rho ^2}{4}+2\gamma +2))
(\log (\mu ^2\rho ^2)+2\gamma-\sum _{s=1}^{2\phi }\frac 1s).
\nonumber\\&&
\label{bttoreg}
\end{eqnarray}

The part of the mass correction to the one-loop action arising from
the term $m^2$ in $M_2$ is hence
\begin{eqnarray}&&
-\frac {m^2\rho ^2}{4}(B_\frac 52-B_\frac 12)
=\frac{m^2\rho ^2}{4}
(\log \frac{m^2\rho^2}{4}+2\gamma +2)(1-3(\log (\mu^2\rho
^2)+2\gamma-\frac {137}{60})
\nonumber\\&&
-\frac{m^2\rho ^2}{4}\frac{77}{60}
\label{zoumzoum}
\end{eqnarray}
where terms not containing $\log \mu $ somewhat inconsequentially
have been retained. The double logarithmic term of (\ref{zoumzoum}) is 
\begin{equation}
-\frac 34m^2\rho ^2\log \frac {m^2\rho^2}{4}\log (\mu ^2\rho ^2)
\end{equation}
in agreement with (\ref{r¿meren}), with $\Lambda $ replaced by $\mu$.
Carrying out in (\ref{acsecond}) a mass renormalization by the
replacement (cf. (\ref{massren})):
\begin{equation}
m^2\rightarrow m^2(1-\frac{g}{32\pi ^2}\log \frac{\mu^2}{\mu '^2})
\label{myssren}
\end{equation}
with $\mu ' $ a new mass parameter, one gets the additional term:
\begin{equation}
\frac 34\log \frac{\mu^2}{\mu '^2}(\log \frac{m^2\rho^2}{4}+2\gamma +2)
\end{equation}
which when added to (\ref{zoumzoum}) replaces 
$\mu ^2$ with $\mu '^2$. Thus it has been shown that the mass
renormalization (\ref{massren}) or (\ref{myssren}) applies for the
whole one-loop mass correction to the action.

The remaining terms of (\ref{maatrixelement}), which are related to
the coupling constant renormalization and the renormalization of the
constraint coefficient, are now inserted into (\ref{logdetto}).  By
means of the relations
\begin{equation}
\sum_{l=0}^{\frac s2-\frac 34}\sum
_{k=1}^{s-\frac 12}\frac{2l+1}{2l+1+k}=\frac 12 (s-\frac 12)^2
\end{equation}
and 
\begin{eqnarray}
&&\sum_{l=0}^{\frac s2-\frac 34}\sum
_{k=1}^{s-\frac 12}\frac{(2l+1)^3}{2l+1+k}
=\frac 13s(s-\frac 12)^2(s+\frac 12)-\frac 18(s^2-\frac 14)^2 
\end{eqnarray}
one obtains the simple expression:
\begin{equation}
-\frac{3m^2\rho ^2}{4}\sum _{s=\frac 72}^\infty 
\frac{s(s^2-\frac 14)}{s^2-\frac{25}{4}}.
\label{romertre}
\end{equation}
Here zeta function
regularization is again applied, with the denominator replaced by the
same quantity to the power $1+\epsilon $ according to the
definition (\ref{Zeta}), converting (\ref{romertre}) into
\begin{equation}
-\frac{3m^2\rho
^2}{4}(Z_\frac 52(\epsilon )+6Z_\frac 52(1+\epsilon)).
\end{equation}
Carrying
out the evaluation by means of (\ref{seksnulsyv}) combined with (\ref{zetminusen}) and
(\ref{zetanen}) one obtains the following additional contribution to
the one-loop action:
\begin{equation}
-\frac{3m^2\rho^2}{4}(-\frac 43+3(\log (\mu ^2\rho ^2)
+2\gamma-\frac{137}{60})).
\label{hundrede}
\end{equation}

The dependence on $\mu$ of (\ref{hundrede}) is removed by a
coupling constant renormalization, where the sum of (\ref{phi4corr})
and (\ref{baresigma}) is multiplied by a factor
\begin{equation}
1-\frac{3g}{32\pi ^2}\log \frac {\mu^2}{\mu '^2}.
\label{ngawbaw}
\end{equation}
This multiplication procedure produces the extra term
\begin{equation}
\frac{9m^2\rho^2}{4}\log \frac {\mu ^2}{\mu '^2}
\end{equation}
that when added added to (\ref{hundrede}) replaces $\mu^2$ with 
$\mu '^2$, demonstrating the consistency of the 
renormalization procedure sketched 
in subsection \ref{Green}, which thus has been confirmed in detail for
the case of an operator constraint. The case of a source constraint is
dealt with in sec.
\ref{SourceC}.

\section{Source constraint} \label{SourceC}
\setcounter{equation}{0}

The renormalization is different with an operator constraint and with a
source constraint since in the latter case the cubic coupling should
not be renormalized and the factor (\ref{persille}) in the path
integral should be replaced by unity. With a source constraint the term
$6\bar{\sigma}_2\phi_{0}$ in $M_2$ is absent, with:
\begin{equation}
6\bar{\sigma}_2\phi_{0}=12m^2u(\log \frac {m^2\rho
^2}{4}+2\gamma +2). \label{sorcerer}
\end{equation}
This induces an
eigenvalue correction to be added to those already determined:
\begin{eqnarray}&&
\delta_{\bar{\sigma}}\lambda_{2, nl}
=-3m^2\rho ^2(\log \frac {m^2\rho ^2}{4}+2\gamma +2)<nl\mid
\mid u^{-1}\mid \mid nl>
\nonumber\\&&
=-3m^2\rho^2(\log
\frac{m^2\rho^2}{4}+2\gamma +2)\frac{2n+4l+3}{2l+1}
\label{ominusen}
\end{eqnarray}
by (\ref{uminusen}).
For $n=0, l=\frac 12$ and  $n=1, l=0$ one finds the following correction
of the translational  and dilatational zero-mode eigenvalues,
respectively
\begin{equation}
\delta_{\bar{\sigma}}\lambda_\mu=-\frac{15}{2}m^2\rho^2(\log
\frac{m^2\rho^2}{4}+2\gamma+2)
\end{equation}
and
\begin{equation}
\delta_{\bar{\sigma}}\lambda_\rho=-15m^2\rho^2(\log
\frac{m^2\rho^2}{4}+2\gamma
+2).
\end{equation}
Both eigenvalues are  nonzero in this case because the source constraint
now also breaks translational invariance.

(\ref{ominusen}) leads to the following correction of the one-loop
action:
\begin{eqnarray}&&
\frac 12\sum_{n+2l>1}(2l+1)^2\frac{\delta_{\bar{\sigma}}
\lambda_{nl}}{\lambda_{nl}}
\nonumber\\&&=-\frac{3m^2\rho^2}{2}(\log
\frac{m^2\rho^2}{4}+2\gamma
+2)\sum_{n+2l>1}\frac{(2l+1)(2n+4l+3)}{(n+2l-1)(n+2l+4)}
\label{zumzum}
\end{eqnarray}
which is divergent and must be regularized along with the
original determinant.

The sum
\begin{equation}
\sum_{n+2l>1}\frac{(2l+1)(2n+4l+3)}{(n+2l-1)(n+2l+4)}
=\sum
_{s=\frac 72}^\infty\frac{s(s^2-\frac
14)}{s^2-\frac{25}{4}}
\label{sumsumsum}
\end{equation}
occurred in (\ref{romertre}) and was evaluated by zeta function
regularization. Using the same procedure here one obtains from
(\ref{zumzum}): 
\begin{equation}
-\frac{3m^2\rho^2}{2}(\log
\frac{m^2\rho^2}{4}+2\gamma
+2)(-\frac 43+3(\log (\mu^2\rho ^2)+2\gamma-\frac{137}{60})).
\label{slawbaw}
\end{equation}
The double logarithmic term of  this result,  with
$\Lambda $ replaced by $\mu$, agrees, apart from the sign, with
(\ref{r¿mertre}), as it should, since the cubic coupling  no more
should be renormalized. However, again the effect of the
renormalization goes beyond the double logarithmic terms. This is seen
from (\ref{baresigma}), which by multiplication with (\ref{ngawbaw})
produces the extra term:
\begin{equation}
-\frac{9m^2\rho^2}{2}\log \frac{\mu ^2}{\mu '^2}(\log
\frac{m^2\rho^2}{4}+2\gamma
+2)
\end{equation} 
with the same dependence on $\mu $ as (\ref{slawbaw}). Thus, with a
source constraint only the mass and coupling constant should be
renormalized, in contrast to the case of an operator constraint, where
a renormalization of the constraint coefficent is also necessary and
is possible because of the factor (\ref{persille}) in the path
integral.

\section{Conclusion}
Our results indicate that constraint terms in the path integral should
be renormalized along with  the Lagrangian. This seems to
suggest that nonrenormalizable constraints are not permitted, thus
leading to a further restriction on the choice of constraint, which in
\cite{NN} was shown to be restricted by the requirement of a finite
classical action.

Only the scalar $\phi ^4$-theory was considered, and thus one should be
cautious by carrying over the result to e.g. the standard model or
supersymmetric gauge theories \cite{Seiberg}, where gauge invariance or
supersymmetry may cause divergences to cancel.

\appendix
\section{Eigenfunctions} \label{Gegencobi}
\setcounter{equation}{0}

Normalized eigenfunctions of (\ref{ayegen})  are:
\begin{eqnarray}&&
\chi _{nl}(u)
=\frac {1}{\Gamma
(2l+2)}\sqrt{\frac{(2n+4l+3)\Gamma (n+4l+3)}{n!
}}
\nonumber\\&&
\hspace{0.1 mm}_2F_1(-n, n+4l+3; 2l+2;u).
\label{chichi}
\end{eqnarray}
Here $\hspace{0.1 mm}_2F_1(-n, n+4l+3; 2l+2;u)$ are hypergeometric
functions (Jacobi polynomials) \cite{Erdelyi}.
The functions $\chi _{nl}(u)$ are normalized with respect to the
integral measure $(u(1-u))^{2l+1}$:
\begin{equation}
\int _0^1du(u(1-u))^{2l+1}\chi _{nl}(u)^2=1.
\label{roderick}
\end{equation}
The Jacobi polynomials are given by a Rodrigues formula and also have  a
convenient series representation:
\begin{eqnarray}
&&\hspace{0.1 mm}_2F_1(-n, n+4l+3; 2l+2;u)
\nonumber\\&&=
\frac{\Gamma (2l+2)}{\Gamma
(n+2l+2)}(u(1-u))^{-2l-1}\frac{d^n}{du^n}(u(1-u))^{n+2l+1}
\nonumber \\&&
=\frac{\Gamma (2l+2)}{\Gamma
(n+4l+3)}\sum
_{k=0}^n(-u)^k\frac{n!}{k!(n-k)!}\frac{\Gamma
(n+4l+3+k)}{\Gamma (2l+2+k)}.
 \label{eeegen}
\end{eqnarray}
  The integral (\ref{roderick}), as well as the
integrals in Appendix
\ref{int}, are evaluated by combination of the Rodrigues formula and
the series representation.

Obviously
\begin{eqnarray}&&
\hspace{0.1 mm}_2F_1(-n, n+4l+3; 2l+2;1-u)
\nonumber\\&&=(-1)^n\hspace{0.1
mm}_2F_1(-n, n+4l+3; 2l+2;u).
\label{symaskine}
\end{eqnarray}
This symmetry reflects the fact that these Jacobi polynomials 
can be expressed in terms of Gegenbauer polynomials $C^{2l+\frac
32}_n(1-2u)$:
\begin{eqnarray}
&&\hspace{0.1 mm}_2F_1(-n, n+4l+3; 2l+2;u)=\frac{n!\Gamma
(4l+3)}{\Gamma (n+4l+3)}C^{2l+\frac 32}_n(1-2u). \label{266}
\end{eqnarray} 
The Jacobi polynomials obey the recursion relation
\begin{eqnarray}
&&(2n+4l+3)(1-2u)\hspace{1 mm}_2F_1(-n, n+4l+3; 2l+2;u)
\nonumber\\&&
=(n+4l+3)
\hspace{1 mm}_2F_1(-n-1, n+4l+4; 2l+2;u)\nonumber\\&&+n\hspace{1
mm}_2F_1(-n+1, n+4l+2; 2l+2;u). \label{retur}
\end{eqnarray}

By differentiation of a Jacobi polynomial is obtained
\begin{eqnarray}&&
\frac{d^{k}}{du^{k}}\vspace{0.6 mm}_2F_1(-n, n+4l+3;
2l+2; u)
\nonumber\\&&
=(-1)^k\frac{n!\Gamma (2l+2)\Gamma
(n+4l+3+k)}{(n-k)!\Gamma (2l+2+k)\Gamma (n+4l+3)} \hspace{1
mm}_2F_1(-n+k, n+4l+3+k; 2l+2+k; u).
\nonumber\\&&
\label{229}
\end{eqnarray}
We also record a number of related relations:
\begin{eqnarray}
&&\frac{d^{n}}{du^{n}}u^{-1}\vspace{0.6 mm}_2F_1(-n, n+4l+3;
2l+2; u)=(-1)^nn!u^{-n-1}, \label{231}
\end{eqnarray}
\begin{eqnarray}
&&\frac{d^{n}}{du^{n}}u^{-2}\vspace{0.6 mm}_2F_1(-n, n+4l+3;
2l+2; u)
\nonumber \\&&
=(-1)^n(n+1)!u^{-n-2}-(-1)^nn!\frac{n(n+4l+3)}{2l+2}u^{-n-1},
\label{232}
\end{eqnarray}
\begin{eqnarray}
&&\frac{d^{n}}{du^{n}}\log u\vspace{0.6 mm}_2F_1(-n, n+4l+3;
2l+2; u)\nonumber\\&&
=(-1)^nn!\frac{\Gamma (2l+2)}{\Gamma (n+4l+3)}\left[(\log u+\sum
_{k=1}^n\frac{1}{k})\frac{\Gamma (2n+4l+3)}{\Gamma (n+2l+2)} 
\right . \nonumber \\&&\left .
-\sum_{k=0}^{n-1}u^{k-n}\frac{1}{n-k}\frac{\Gamma(n+4l+3+k)}{\Gamma
(2l+2+k)}\right ], \label{233}
\end{eqnarray}
\begin{eqnarray}
&&\frac{d^{n-1}}{du^{n-1}}\log u\vspace{1 mm} _2F_1(-n, n+4l+3;
2l+2; u)
\nonumber \\&&
=(-1)^nn!\frac{\Gamma (2l+2)}{\Gamma (n+4l+3)}\left[(\log u
(u-\frac 12)+u\sum _{k=2}^n\frac{1}{k}-\frac 12\sum
_{k=1}^{n-1}\frac 1k)\frac{\Gamma (2n+4l+3)}{\Gamma (n+2l+2)} 
\right . \nonumber \\&&\left .
+\sum_{k=0}^{n-2}u^{k-n+1}\frac{1}{(n-k)(n-k-1)}\frac{\Gamma(n+4l+3+k)}{\Gamma
(2l+2+k)}\right ] \label{2333}
\end{eqnarray}
and
\begin{eqnarray}
&&\frac{d^{n+1}}{du^{n+1}}\log u\vspace{1 mm} _2F_1(-n, n+4l+3;
2l+2; u)
\nonumber \\&&
=(-1)^{n}n!\frac{\Gamma (2l+2)}{\Gamma (n+4l+3)}
\sum_{k=0}^{n}u^{k-n-1}\frac{\Gamma(n+4l+3+k)}{\Gamma
(2l+2+k)}. \label{23333}
\end{eqnarray}

\section{Integrals} \label{int}
\setcounter{equation}{0}

Matrix elements involving the functions $\chi _{nl}(u)$
are evaluated here, with the notation:
\begin{equation}
<nl\mid \mid F(u)\mid \mid nl>
=\int _0^1du (u(1-u))^{2l+1}F(u)\chi ^2_{nl}(u).
\label{chichichi}
\end{equation}

By means of (\ref{231}) one finds: 
\begin{equation}
<nl\mid \mid u^{-1}\mid \mid nl>=\frac{2n+4l+3}{2l+1} 
\label{uminusen}
\end{equation}
 and similarly by means of (\ref{233}):
\begin{eqnarray}
&&<nl\mid \mid u^{-2}\mid \mid nl>
=\frac{(n+2l+1)(n+2l+2)(2n+4l+3)}{2l(l+1)(2l+1)}.
\label{enoveruianden}    
\end{eqnarray}
This expression is only well-defined for
$l\neq 0$. For $l=0$ also boundary terms arise through partial
integrations.  Thus  a lower cutoff in the $u$-integration $u_{\rm
min}\simeq \frac {\rho^2}{R^2}$ is introduced corresponding to the
integral in coordinate space being restricted to a large sphere of
radius $R$. Using also 
\begin{equation}
\int _0^1duu^{p-1}(1-u)^{q-1}\log u=-B(p,q)
\sum_{ k=0}^{q-1}\frac{1}{p+k} \label{intlog}
\end{equation}
with $B(p,q)$ Euler's beta function, one finds the total matrix element
in this case:
\begin{eqnarray}&&
<n0\mid \mid u^{-2}\mid \mid n0>
\nonumber\\&&=2n+3
+(2n+3)(n+1)(n+2)(\log \frac{R^2}{\rho
^2}-\frac 32-2\sum _{k=1}^n\frac{1}{k+1}).
\label{l0}
\end{eqnarray}
Also the result
\begin{equation}
<nl\mid \mid \log u \mid \mid nl>
=-\sum _{k=0}^{n+2l+1}\frac{1}{n+2l+2+k}
-\sum _{k=0}^{n-1}\frac{1}{n+4l+3+k}
\label{enelog}
\end{equation}
follows from (\ref{233}) and (\ref{intlog}).

A more complicated matrix element is
\begin{eqnarray}
&&<nl\mid \mid \frac{\log u }{u(1-u)}\mid \mid nl>
\nonumber \\&&
=(-1)^n\frac{(2n+4l+3)
\Gamma (n+4l+3)}{\Gamma (2l+2)\Gamma (n+2l+2)}
\int_0^1 du (u(1-u))^{n+2l+1}
\nonumber \\&&
 \sum
_{k=0}^n\frac{1}{k!(n-k)!}(\frac{d^{k}}{du^{k}}\frac{\log
u}{u(1-u)})
\frac{d^{n-k}}{du^{n-k}}\vspace{0.6 mm}_2F_1(-n, n+4l+3; 2l+2;
u) )). 
\nonumber\\&&
\label{277}
\end{eqnarray}
The following useful relation is readily proved by induction:
\begin{eqnarray}
&&\frac{d^n}{du^n}\frac{\log
u}{u(1-u)}
=n!\log
u(\frac{(-1)^n}{u^{n+1}}+\frac{1}{(1-u)^{n+1}}) 
\nonumber\\&&
+n!(-1)^n \sum
_{k=1}^n\frac{1}{k}\frac{1}{u^{k}}\sum
_{s=1}^{n-k+1}\frac{(-1)^s}{u^{n+2-k-s}(1-u)^s}.
\label{loguha}
\end{eqnarray}
The  contribution from the part of  (\ref{277}) with a logarithmic
integrand is:
\begin{eqnarray}
&&(-1)^n\frac{(2n+4l+3)
\Gamma(n+4l+3)}{n!\Gamma (2l+2)\Gamma(n+2l+2)}\int_0^1 du
(u(1-u))^{n+2l+1}\log u \frac{d^n}{du^n}(u(1-u))^{-1}
\nonumber \\&&
\hspace{1 mm}_2F_1(-n, n+4l+3; 2l+2;u))
\nonumber \\&&
=-\frac{2n+4l+3}{2l+1}(\sum_{k=0}^{n+2l+1}\frac{1}{2l+1+k}
+\sum_{k=0}^{2l}\frac{1}{n+2l+2+k}).
\label{logto} 
\end{eqnarray}
by (\ref{231}) and (\ref{intlog}).
The nonlogarithmic part of the integrand of (\ref{277}) yields by
(\ref{229}) and (\ref{loguha})
\begin{equation}
-\frac{2n+4l+3}{2l+1}\sum
_{k=1}^{n}\frac{1}{2l+k+1}
\label{nonlog}
\end{equation}
where also the following algebraic identities were used:
\begin{eqnarray}&&
\sum_{p=0}(-1)^p\frac{k!}{p!(k-p)!}\frac{\Gamma
(n+2l-k+p+s)}{\Gamma (n+2l+2-k+p)}(2n+4l+2-k+p)
\nonumber\\&&=
\left\{\begin{array}{c}
            \frac{(n+2l-k)!k!}{(n+2l)!}\hspace{1 mm}{\rm
for}\hspace{1 mm}s=1\\0\hspace{1 mm}{\rm for}\hspace{1 mm}s\geq 2
            \end{array}  \right .   
\end{eqnarray}
and
\begin{equation}
\sum
_{k=1}^{n}\frac{n!(n+2l-k)!}{(n-k)!(n+2l+1)!}\sum
_{r=1}^k\frac 1r=\frac{1}{2l+1}\sum _{k=1}^n\frac{1}{2l+1+k}.
\label{noonlog}
\end{equation}
Adding  (\ref{logto}) and (\ref{nonlog}) one obtains:
\begin{equation}
<nl\mid \mid \frac{\log u }{u(1-u)}\mid \mid
nl>=-\frac{2n+4l+3}{2l+1}(2\sum
_{k=1}^{n+2l+1}\frac{1}{2l+1+k}+\frac{1}{2l+1}).
\label{hunafloi}
\end{equation}

The last matrix element needed is
\begin{eqnarray}
&&<nl\mid \mid (u-\frac 12)\Phi (\frac{1-u}{u})\mid \mid
nl>\nonumber
\\&&
=
-(-1)^n\frac{\Gamma (n+4l+3)}{2n!\Gamma (2l+2)\Gamma (n+2l+2)}
\int_0^1 du (u(1-u))^{n+2l+1}\frac{d^{n}}{du^{n}}\Phi
(\frac{1-u}{u})
\nonumber\\&&
((n+4l+3)
\vspace{0.6 mm}_2F_1(-n-1, n+4l+4; 2l+2; u)
\nonumber\\&&
+n\vspace{0.6 mm}_2F_1(-n+1, n+4l+2; 2l+2; u))
\label{Finally}
\end{eqnarray}
where the recursion relation (\ref{retur}) was used.

(\ref{Finally}) has the Spence function part:
\begin{eqnarray}
&&(-1)^{n+1}\frac{
\Gamma (n+4l+4)}{2n!\Gamma (2l+2)\Gamma (n+2l+2)}
\int_0^1 du (u(1-u))^{n+2l+1}\Phi (\frac{1-u}{u})
\nonumber \\&&\frac{d^{n}}{du^{n}}
\vspace{0.6
mm}_2F_1(-n-1, n+4l+4; 2l+2; u))
\nonumber \\&&
=-\frac{n+1}{2n+4l+4}\sum
_{k=0}^{n+2l+1}\frac{1}{n+2l+2+k}
\label{Spyncto}
\end{eqnarray}
evaluated by means of (\ref{229}) and 
\begin{equation}
\int _0^1 du u^{p-1}(1-u)^{q-1}\Phi (\frac{1-u}{u})=B(p,q)(\zeta
(2,p)+\sum _{r=1}^{q-1}\frac 1r\sum _{s=0}^{r-1}\frac{1}{p+s})
\label{Folmer}
\end{equation}
where
$\zeta(2,p)=\sum _{s=p}^\infty \frac{1}{s^2}$
is a generalized zeta function.

The logarithmic term of (\ref{Finally}) is by (\ref{symaskine}):
\begin{eqnarray}&&
(-1)^{n+1}\frac{
\Gamma (n+4l+4)}{2n!\Gamma (2l+2)\Gamma (n+2l+2)}
\int_0^1 du (u(1-u))^{n+2l+1}\log (u(1-u))
\nonumber\\&&
(\frac{d^{n}}{du^{n}}\log u
\vspace{0.6 mm}_2F_1(-n-1, n+4l+4; 2l+2; u)
\nonumber\\&&-\log u
\frac{d^{n}}{du^{n}}
\vspace{0.6 mm}_2F_1(-n-1, n+4l+4; 2l+2; u))
\nonumber\\&&
+(-1)^{n-1}\frac{
\Gamma (n+4l+3)}{2(n-1)!\Gamma (2l+2)\Gamma (n+2l+2)}
\int_0^1 du (u(1-u))^{n+2l+1}\log (u(1-u))
\nonumber\\&&
(\frac{d^{n}}{du^{n}}\log u
\vspace{0.6 mm}_2F_1(-n+1, n+4l+2; 2l+2; u)
\nonumber\\&&-\log u
\frac{d^{n}}{du^{n}}
\vspace{0.6 mm}_2F_1(-n+1, n+4l+2; 2l+2; u)).
 \label{entlog}
\end{eqnarray}
Here each  term  is evaluated separately by means
of (\ref{2333}), (\ref{23333}) and (\ref{intlog}). The first term of
(\ref{entlog}) is 
\begin{eqnarray}&&
\frac{n+1}{2n+4l+4}\sum
_{k=0}^{n-1}(\frac{1}{n+4l+4+k}-\frac{1}{2l+2+k})
\nonumber \\&&
+\frac 12\sum _{k=0}^{n-1}\frac{1}{2l+2+k}-\sum
_{k=0}^{n+2l+1}\frac{1}{n+2l+2+k}+\sum
_{k=0}^{2l+1}\frac{1}{n+2l+2+k}
\nonumber\\&&
\label{fire}
\end{eqnarray}
and the
second term of (\ref{entlog}) is
\begin{eqnarray}&&
-\frac 12 \sum
_{k=0}^{n}\frac{1}{2l+1+k}+\frac{n+4l+2}{2n+4l+2}\sum
_{k=0}^{n-1}(\frac{1}{2l+1+k}-\frac{1}{n+4l+2+k})+\frac{n+4l+2}{4(n+2l+1)^2}
\nonumber\\&&
-\sum
_{q=0}^{2l}\frac{1}{n+2l+2+q}-1+\frac{n+4l+2}{2n+4l+2}
+\frac{n+4l+2}{2n+4l+2}\sum _{q=0}^{n+2l}\frac{1}{n+2l+2+q}
\label{seks}
\end{eqnarray}
by some algebraic manipulations.

Finally  (\ref{Finally}) has the following contribution from the
nonlogarithmic part of the integrand according to (\ref{229}) and
(\ref{loguha}):
\begin{eqnarray}&&
(-1)^n\frac{\Gamma (n+4l+3)}{2\Gamma (2l+2)\Gamma (n+2l+2)}
\int_0^1 du
(u(1-u))^{n+2l+1}\sum_{k=1}^n\frac{(-1)^k}{k(n-k)!}
\nonumber\\&&\sum
_{r=1}^{k-1}\frac{1}{r}\sum
_{s=1}^{k-r}\frac{(-1)^{s}}{u^{k+1-s}(1-u)^s}
\frac{d^{n-k}}{du^{n-k}}((n+4l+3)
\vspace{0.6 mm}_2F_1(-n-1, n+4l+4; 2l+2; u)
\nonumber\\&&
+n\vspace{0.6 mm}_2F_1(-n+1, n+4l+2; 2l+2; u))
\nonumber\\&&
=\frac{1}{2\Gamma (n+2l+2)}
\sum _{k=1}^n\frac{n!}{k(n-k)!}\sum_{r=1}^{k-1}\frac{1}{r}
\sum_{s=1}^{k-r}(-1)^s\Gamma(n+2l+2-s)
\nonumber \\&&
((n+1)\sum _{p=0}^{k+1}(-1)^{p}\frac{1}{p!(k+1-p)!}
\frac{\Gamma (n+2l+1-k+p+s)}{\Gamma (n-k+2l+2+p)}
(2n-k+4l+3+p)
\nonumber \\&&
+(n+4l+2)\sum _{p=0}^{k-1}(-1)^{p}\frac{1}{p!(k-1-p)!}
\frac{\Gamma (n+2l+1-k+p+s)}{\Gamma (n-k+2l+2+p)}
\frac{1}{2n-k+4l+2+p}).
\nonumber\\&& \label{285}
\end{eqnarray}
Here the following algebraic identities are used:
\begin{eqnarray}&&
\sum_{p=0}^{k+1}(-1)^p\frac{(k+1)!}{p!(k+1-p)!}
\frac{\Gamma
(n+2l+1-k+p+s)}{\Gamma (n+2l+2-k+p)}(2n-k+4l+3+p)=0
\nonumber\\&&
\label{glenquin}
\end{eqnarray}
and
\begin{eqnarray}
&&\sum
_{p=0}^{k-1}(-1)^{p}\frac{1}{p!(k-1-p)!}
\frac{\Gamma (n+2l+1-k+p+s)}{\Gamma
(n-k+2l+2+p)}\frac{1}{2n-k+4l+2+p}
\nonumber \\&&
=(-1)^{s-1}\frac{\Gamma(n+2l+1)}{\Gamma(n+2l+2-s)}
\frac{\Gamma(2n+4l+2-k)}
{\Gamma(2n+4l+2)}.
\end{eqnarray}
Thus (\ref{285}) is:
\begin{eqnarray}&&
-\frac{n+4l+2}{2(n+2l+1)(2n+4l+1)!}\sum
_{k=1}^n\frac{n!(2n+4l+1-k)!}{k(n-k)!}\sum_{r=1}^{k-1}\frac{1}{r}
(k-r)
\nonumber\\&&
=-\sum_{k=1}^n\frac{1}{n+4l+2+k}+1-\frac{n+4l+2}{2n+4l+2}
\label{288+291}
\end{eqnarray}
by (\ref{noonlog}) and the identity 
\begin{equation}
\sum _{k=1}^n\frac{n!(2n+4l+1-k)!}{(n-k)!(2n+4l+2)!}=\frac{1}{n+4l+2}.
\end{equation}

Adding (\ref{Spyncto}), (\ref{fire}), (\ref{seks}) and (\ref{288+291})
one finally obtains:
\begin{eqnarray}&&
<nl\mid \mid (u-\frac 12)\Phi (\frac{1-u}{u})\mid \mid
nl>
\nonumber\\&&
=-\frac{n+1}{2n+4l+4}\sum
_{k=0}^{n+2l+1}\frac{1}{2l+2+k}-\sum
_{k=0}^{n-1}\frac{1}{n+4l+4+k}+ \frac{1}{4l+2}-\frac{1}{n+4l+2}
\nonumber\\&&+\sum
_{k=0}^{n-1}\frac{1}{2l+2+k}
-\frac{n}{2n+4l+2}\sum
_{k=0}^{n+2l}\frac{1}{2l+1+k}-
\sum _{k=0}^{n-1}\frac{1}{n+4l+3+k}.
\label{looklook}
\end{eqnarray}

\section{Zeta function regularization}
\label{appc}
\setcounter{equation}{0}

\subsection{The zeta function}

The Riemann zeta function is \cite{Erdelyi}
\begin{equation}
\zeta (s)=\sum _{k=1}^\infty k^{-s}=\frac{1}{\Gamma (s)}\int
_0^\infty dt
\frac{t^{s-1}}{e^{t}-1}.
\label{Birn}
\end{equation}
The zeta function behaves near 
$\epsilon =0$  according to:
\begin{equation}
\epsilon \zeta
(1+2\epsilon)\simeq 
\frac 12+\gamma \epsilon \label{zetlog}
\end{equation}
where $\gamma $ is Euler's constant. The zeta function as defined in
(\ref{Birn}) is only well-defined for
$s>1$, but can be analytically continued to the whole complex plane.

The generalized zeta function is with $a>0$, $s>1$:
\begin{equation}
\zeta (s, a)=\sum _{k=0}^\infty (k+a)^{-s}=\frac{1}{\Gamma
(s)}\int _0^\infty t^{s-1} dt \frac{e^{t(1-a)}}{e^{t}-1}.
\label{zetaaet}
\end{equation}
It also also has an analytic continuation in the variable $s$ and obeys
for all values of
$s$ the functional equation
\begin{equation}
\zeta (s, a)=\zeta (s, a+m)+\sum _{k=0}^{m-1} (k+a)^{-s}. \label{Funk}
\end{equation}

\subsection{Zeta function regularization} \label{ZZeta}
For a general self-adjoint  operator $\Delta $ with
eigenvalues $\lambda $  a
generalized  zeta  function is formed \cite{Hawking}, \cite{Elizalde}: 
\begin{equation}
\zeta _\Delta (\epsilon)=\sum \lambda ^{-\epsilon}
\label{defzeta}
\end{equation}
and  in the absence of zero modes the determinant of $\Delta$ is
defined by: 
\begin{equation}
\log\det \Delta=-\lim _{\epsilon \rightarrow 0}\frac
{\partial}{\partial \epsilon}\mu ^{2\epsilon}\zeta _\Delta
(\epsilon)=-\log (\mu ^2)\zeta_\Delta (0)-\zeta_\Delta
'(0).
\label{undet} \label{zetadet}
\end{equation}
with $\mu $ an arbitrary mass scale (supposing $\Delta $ has
dimension mass squared).

When $\Delta $ is perturbed the eigenvalue $\lambda $ is changed by an
amount $\delta \lambda$. Hence the 
zeta function is changed by the amount 
\begin{equation}
\delta \zeta_\Delta(\epsilon)=-\epsilon\sum \delta \lambda
\lambda ^{-1-\epsilon}
\label{D105}
\end{equation}
and 
\begin{equation}
\delta \log \det \Delta=\log (\mu^2)\lim _{\epsilon
\rightarrow 0}\epsilon\sum \delta \lambda \lambda
^{-1-\epsilon}+\lim _{\epsilon \rightarrow 0}\frac{\partial
}{\partial \epsilon}
\epsilon\sum \delta \lambda \lambda
^{-1-\epsilon}. \label{seksnulsyv}
\end{equation}

\subsection {The function $Z_\phi(\epsilon )$}

Zeta function regularization involves the function $Z_\phi(\epsilon )$
defined in (\ref{Zeta}); it has a useful representation in terms of a
Feynman parameter
$\alpha
$:
\begin{eqnarray}&&
Z_\phi(\epsilon )
=\frac{1}{\Gamma ^2(\epsilon )}\int
_0^\infty dtt^{2\epsilon-1}\int _0^1d\alpha (\alpha (1-\alpha
))^{\epsilon -1}\frac{e^{-2t\phi(1-\alpha)}}{e^t-1}
(\frac{2\epsilon-1}{t}-\phi(1-2\alpha))).
\nonumber\\&&
\label{eensekssyv}
\end{eqnarray}
By a power series expansion of the exponential and use of
(\ref{Birn}) one expresses $Z_\phi(\epsilon )$ as an infinite sum of
Riemann zeta functions: 
\begin{eqnarray}&&
Z_\phi(\epsilon)
=\zeta (2\epsilon-1)-\phi (2\epsilon-1)\zeta (2\epsilon
)
\nonumber\\&&
+\sum _{n=2}^\infty \frac{(-2\phi )^n}{(n-2)!}(\frac{(\epsilon
+n-1)(2\epsilon-1)}{n(n-1)}+\frac 12)\frac{\Gamma (\epsilon
+n-1)}{\Gamma (\epsilon )}\frac{1}{2\epsilon+n-1}\zeta(2\epsilon +n-1)
\nonumber\\&&
\label{zetsum}
\end{eqnarray}
whence
\begin{equation}
Z_\phi(0)=-\frac{1}{12}-\frac{\phi }{2},
Z_\phi(-1)=\frac{1}{120}-\frac{\phi ^2}{6}. \label{zetminusen}
\end{equation}

From (\ref{zetsum})  and (\ref{Birn})-(\ref{Funk}) follows for
$\epsilon \simeq 0$:
\begin{equation}
\epsilon Z_\phi(1+\epsilon
)\simeq \frac 12+\epsilon (\gamma-\frac 12\sum _{s=1}^{2\phi }\frac
1s).
\label{zetanen}
\end{equation}

\end{document}